\documentclass[aps,twocolumn]{revtex4}
\usepackage{bm}
\usepackage[a4paper, margin=1in]{geometry}
\usepackage{amsmath}
\usepackage{upgreek}
\usepackage{nameref,hyperref,cleveref}
\usepackage[english]{babel}
\usepackage[utf8]{inputenc}
\DeclareUnicodeCharacter{2212}{-}
\usepackage{lipsum}
\usepackage{graphicx}
\usepackage[colorinlistoftodos, color=green!40, prependcaption]{todonotes}
\begin{document}
\title{ A study of pure multi-strange hadrons production in Pb+Pb collisions at LHC energies using HYDJET++ model}
\author{Gauri Devi$^{1}$\footnote{gauri.devi13@bhu.ac.in} and  B. K. Singh$^{1}$\footnote{bksingh@bhu.ac.in, director@iiitdmj.ac.in(Corresponding author)}$^{,2}$}
 
\affiliation{$^{1}$ Department of Physics, Institute of Science, Banaras Hindu University (BHU), Varanasi 221005, India  \\ $^{2}$ Discipline of Natural Sciences, PDPM Indian Institute of Information Technology Design \& Manufacturing, Jabalpur 482005, India }

\begin{abstract}
For the present work, we have used the HYDJET++ model to explore the production of pure multi$-$strange hadrons in Pb+Pb collisions at $\sqrt{s_{NN}}$= 2.76 TeV and $\sqrt{s_{NN}}$= 5.02 TeV collision energies, respectively. We have performed simulation to investigate transverse momentum ($p_{T}$) spectra and elliptic flow ($v_{2}$) for $\phi-$meson and $\Omega-$baryons, and compared the results with ALICE experimental data as well as predictions from various phenomenological models across different centrality classes. Furthermore, we have calculated the nuclear modification factors ($R_{AA}$ and $R_{CP}$), which provide a perception of jet quenching phenomena. Hence, our findings enable the study of the energy and system dependence of $\phi$ and $\Omega$ hadrons production over a wide range of ultra-relativistic collision energies. We also present the particle ratios ( $\Omega/\phi$, $\overline{\Omega}^{+}/\Omega^{-}$, $\Omega/\pi$, and $\phi/\pi$ ), offering insights into the strangeness enhancement and chemical properties of the medium at both LHC collision energies. Additionally, we explored the predictions for $\Omega$-baryon in Pb+Pb collisions at$\sqrt{s_{NN}}$= 5.02 TeV, focusing on $R_{AA}$, $R_{CP}$, and particle ratios within the HYDJET++ framework, offering insights into future measurements and particle dynamics in high-energy heavy-ion collisions.
\end{abstract}

\maketitle
\section*{Keywords:}
 Quark Gluon Plasma, multi$-$strange Hadrons, Jet-quenching, Decoupling, Multiple Kinetic Freeze-out Temperature, HYDJET++.
 
\section{Introduction}
One of the most important theoretical tasks for quark gluon plasma (QGP) formation in heavy ion collisions at ultra-relativistic energies is to separate or account for the influence of the QGP phase and hadron gas expansion from true signals of phase transformation. To solve this task, one needs a reliable theoretical and experimental description of hadron gas properties emitted from the hypersurfaces of chemical and thermal freeze-out~\cite{STAR:2022tfp, Devi:2024uis,Waqas:2021rmb, Waqas:2021bsy,ToporPop:2007hb}. \\
~~Both $\phi$ and $\Omega$ particles decouple from the surrounding system earlier than non-strange hadrons due to their smaller hadronic interaction cross-sections, higher mass, and lower interaction rates~\cite{STAR:2015gge,Bugaev:2016voh}. This early decoupling refers to the process by which particles cease interacting strongly with the surrounding medium, providing more direct information about the chemical/thermal freeze-out stage which is expected near the quark-hadron transition temperature ~\cite{STAR:2015gge,Bugaev:2016voh}. These particles experience little to no distortion due to hadronic rescattering and are less affected by the pressure generated by the hadronic matter in the later stages of the reaction~\cite{NA49:2004irs}.\\
As a result, $\phi$ and $\Omega$ 
particles provide a unique tool for probing the transition from partonic to hadronic dynamics~\cite{STAR:2015vvs}. Additionally, the study of $\phi$ strongly suggests a baryon versus meson dynamic and strangeness enhancement, as opposed to a simple mass dependence, as would be the case for radial flow developed during the hadronic phase~\cite{PHENIX:2013kod}. It arouses a great interest in QGP fireball formation as a function of energy~\cite{STAR:2019bjj, Ye:2017ooc} and system size ~\cite{Devi:2024dvp, ALICE:2021lsv} at ultra-relativistic nuclear collisions.\\
~~The mechanism of strangeness production in ultra-relativistic heavy-ion collisions can be explored by analyzing the particle spectra,  $\langle p_{T}\rangle$, and anisotropic flow. The shape of $p_{T}$ spectra varies for different particles, indicating that heavier particles gain more momentum at a given velocity when thermally produced from the QGP fluid due to the collective flow in the expanding medium and also provides information about initial energy density. Mean $\langle p_T\rangle$ of particles produced in a collision is related to the initial transverse size of the fireball, where a higher $\langle p_{T}\rangle$ value is indicating a smaller initial area~\cite{STAR:2019bjj}. These measurements reveal possible flavour dependencies and help to distinguish between soft and hard processes based on quark content~\cite{PHENIX:2004spo, Singh:2023bzm,Feng:2022sdh}. Likely, the ratio of the $p_T$-spectra at $\sqrt{s_{NN}}=$ 5.02 TeV and $\sqrt{s_{NN}}=$ 2.76 TeV as a function of $p_T$ may help to understand the strangeness enhancement and medium effect through nuclear modification factors ($R_{AA}$ and $R_{CP}$)~\cite{ALICE:2021ptz}. The enhanced multiplicity of strange particles in the QGP compared to the hadron gas (HG) phase can be referred to the increased reaction volume of large nuclei, as described in Braun-Munzinger's work~\cite{Braun-Munzinger:2003htr}.\\
In addition to the volume effect, the under-saturation of strange particle phase space introduces a strangeness suppression factor ($\gamma_s$)~\cite{Braun-Munzinger:2003htr}. The deviation of the strange particle yields from a hadron gas in full equilibrium was quantified by $\gamma_s$. This suppression is weaker in heavy-ion collisions and fitting hadron multiplicities in full-phase space still require consideration of $\gamma_s$. One reasonable explanation is that the total strangeness available for hadronization is set during a prehadronic stage of the collision. Thus, a change in $\gamma_s$ between $p+p$ and $A+A$ collisions may reflect differences in the initial conditions of the different fireballs~\cite{NA49:2008goy}.\\
~~Furthermore, anisotropic flow ($v_{2}$)~\cite{ALICE:2018yph, ALICE:2014wao, STAR:2003wqp, Devi:2023wih} as a function of $p_T$ and centrality provides information about the pressure gradients, the effective degrees of freedom, the degree of thermalization, and the equation of state of the matter created at the early stage in non-central collisions. At low $p_{T}$, flow shows a hydrodynamic expansion in the system and mass orders of $v_{2}$. At intermediate to high $p_T$, there is a noticeable separation in the baryon-meson spectra as a function of $p_T$. However, no distinction is observed between baryons and mesons as a function of centrality, supporting the NCQ scaling~\cite{ALICE:2014wao, Devi:2023wih}. $v_{2}$ of baryons is larger than $v_{2}$ of mesons at intermediate to high $p_T$ gives additional insight into hadronization through quark coalescence ~\cite{ALICE:2014wao} and includes interactions of jet fragments with bulk matter. In this $p_T$ range, a non-zero value of $v_{2}$ is generated when hard partons propagating through the system lose energy via radiative and collisional loss processes~\cite{ALICE:2021ptz,Ortiz:2017cul}. This allows us to safely assume that both jet quenching and $v_{2}$ at high $p_{T}$ are dominated by energy loss~\cite{Christiansen:2013hya}. The influence of flow would result in an underestimation (overestimation) of the quenching contribution in (out) of the plane during heavy ion collisions~\cite{ALICE:2018yph, ALICE:2014wao, Devi:2023wih,Christiansen:2013hya}.\\
In this paper, we employ the HYDJET++ model~\cite{Lokhtin:2008xi, Lokhtin:2009hs, Lokhtin:2005px} to describe the behaviour of (multi$-$) strange hadrons in heavy-ion collisions and also compare its results with predictions from other well-known simulation models such as AMPT, VISHNU, EPOS, $\text{HIJING}/B\overline{B}$, and the Krakow model~\cite{Zhu:2015dfa, Zhu:2016qiv,Lin:2021mdn, Lin:2004en,He:2017tla,Pierog:2013ria,Bozek:2012qs, ToporPop:2011wk, Werner:2007bf,ToporPop:2010qz,Devi:2024uis}, each offering different approaches to simulating particle production and dynamics in heavy-ion collisions.\\
AMPT~\cite{He:2017tla,Lin:2021mdn} describes heavy-ion collisions by incorporating fluctuating initial conditions, two-body elastic parton scatterings, hadronization, and hadronic interactions. The default version of the AMPT model, which includes only minijet partons in the parton cascade and employs the Lund string fragmentation~\cite{Cimerman:2017lmm} for hadronization, can well describe the rapidity distributions and $p_T$-spectra of identified particles in heavy-ion collisions from SPS to LHC energies. However, it significantly underestimates the elliptic flow at RHIC and (multi$-$) strange spectra at RHIC and LHC. To address this, the AMPT model with String Melting (SM) incorporates a quark coalescence mechanism for hadronization~\cite{Lin:2004en}. While the improved coalescence mechanism in the AMPT-SM model enhances the better description of strange particle ratios, anisotropic flow, it still underpredicts the strange yield~\cite{Lin:2021mdn}. This deviation is likely due to limitations in the coalescence model's treatment of strange quark production and dynamics, as well as the initial conditions provided by HIJING that further discussed in the HIJING description, which may not fully capture the abundance of strange quarks necessary to match experimental data.\\ In the same context, the EPOS and Krakow models are based on the core-corona concept~\cite{Singh:2023bzm}. In EPOS, this separation depends on the initial energy density-areas with high energy density are called the core, while low-density areas
are the corona~\cite{Werner:2007bf}. This helps in explaining how particle production changes with collision centrality. In central collisions, the core dominates and plays a major role in producing heavier and multi$-$strange particles because of the dense environment and higher energy, leading to particles with larger transverse momentum ($p_T$). In contrast, the corona contributes more in peripheral collisions, where the density is lower.\\ 
The Krakow~\cite{Bozek:2012qs} is a $(3+1)-$dimensional hydrodynamic model that includes essential parameters like bulk and shear viscosities. In this model, nonequilibrium corrections can be described through the bulk viscosity coefficient in the expanding fireball, increasing the effective chemical freeze$-$out temperature~\cite{Bozek:2012qs}. However, this effect is not strong enough to reproduce the observed yield of heavy hadrons. This agreement could be improved using different equilibration rates and different freeze-out temperatures for different particle species~\cite{Bozek:2012qs}.\\ Another important model in this context is VISHNU, which combines VISH$2+1$ for QGP fluid expansion and UrQMD for hadron resonance gas evolution. However, the UrQMD module only includes strong resonance decays and neglects weak decays and baryon$-$antibaryon ($B−\overline{B}$) annihilations. This omission leads to a significant reduction in the yields of strange and multi$-$strange baryons about $30\%$ for $\Lambda$ and $20\%$ for $\Xi$ and 
$\Omega$ in the most central Pb+Pb collisions~\cite{Zhu:2015dfa,He:2017tla}. 
Additionally, the differences between the calculated $p_T$ spectra and the 
experimental data, along with the mismatch in mass ordering among $p$, $\Lambda$, and $\Xi$, suggesting that the results cannot be fully explained by assuming a single chemical freeze$-$out temperature~\cite{Zhu:2015dfa,He:2017tla}.\\
Following this, $\text{HIJING}/B\overline{B}$ model~\cite{ToporPop:2011wk, ToporPop:2010qz} is an upgraded version of regular HIJING within the new parameters, strong colour field(k) and  various suppression factors
(\emph{i.e.}, diquark, strangeness, suppression of spin 1 diquarks relative to spin 0 ones, and normal) based on $\text{HIJING}/B\overline{B}v1.10$ and HIJING model~\cite{ToporPop:2007hb}.
These modifications enhance the production of (multi-) strange particles compared to the regular HIJING model but still fall short of matching experimental data. 
Similar to HYDJET++, these changes have little impact on bulk particle yields, allowing the model to describe both non$-$strange and multi-strange hadrons simultaneously.
Additionally, it supports the theoretically and experimentally
~\cite{Waqas:2021rmb, Waqas:2021bsy, Cleymans:2004pp, Cleymans:2006xj} observed
early thermal freeze-out of multi-strange hadrons through fluctuations 
of the transient strong colour-field~\cite{ToporPop:2007hb}.\\
The paper is organized as follows: Section~\ref{Model formalism} briefly describes the HYDJET++ model, which supports the early thermal and chemical freezeout of multi$-$strange particles. Section \ref{A} highlights the measurements of $\phi$ and $\Omega$ hadrons spectra obtained by the HYDJET++ model. These results are compared to available ALICE experiment data and other available models AMPT, VISHNU, and EPOS results, focusing on collisions at LHC energies ~\cite{ALICE:2013xmt, ALICE:2014jbq} as a function of $p_T$. Section~\ref{B} discusses the factors, $R_{CP}$ and $R_{AA}$. Section~\ref{C} explores the $\Omega/\phi$, $\phi/\pi^{-}+\pi^{+}$, $\overline{\Omega}^{+}+\Omega^{-}$/$\pi^{-}+\pi^{+}$, and $\overline{\Omega}^{+}/\Omega^{-}$ ratios with available experimental data and other Krakow and HIJING/$B\overline{B}$ model results~\cite{Bozek:2012qs, ToporPop:2011wk, ToporPop:2010qz,Devi:2024uis}. Section~\ref{D} evaluates the elliptic flow ($v_2$) as a function of $ p_T$, $\langle p_T\rangle$, and $\langle v_2\rangle$ as a function of $N_{part}$. The measured mean $\langle p_T\rangle$ also shows the variation with energies and heavier mass/quark content, which provides more information about the thermal/chemical equilibrium and strong radial flow. 
Section~\ref{E} discusses charged particles mass ordering phenomena at ($10-20\%$) centrality interval. Finally, section~\ref{conclusions} summarizes our findings. 
\section{Model Formalism}\label{Model formalism}
HYDJET++ model is a widely used hybrid Monte Carlo event generator that successfully simulates a large number of observables measured in relativistic heavy-ion collisions~\cite{Lokhtin:2005px, Lokhtin:2009hs} at RHIC as well as LHC energies. The HYDJET++ simulation is comprised of two independent processes: one simulating soft, hydro process (low momentum regime) which is based on preset relativistic hydrodynamics parametrization of the freeze-out hypersurfaces
by FAST MC~\cite{ Amelin:2007ic} generator and the other simulates hard component where energetic partons lose energy in the medium. 
 The soft component simulates a thermalized hadronic system generated on freeze-out hypersurfaces defined by parameterized relativistic hydrodynamics with specified freeze-out conditions~\cite{Amelin:2007ic,Crkovska:2016flo}. The effective thermal volume of the fireball, calculated for each event, determines the average hadron multiplicities at freeze-out and is proportional to the number of wounded nucleons at a given collision centrality, as provided by the Glauber model~\cite{Piasecki:2023hoo}. Final-state interactions are limited to two- and three-body decays of resonances, which are taken from an extensive table containing over 360 mesons and baryons states~\cite{Torrieri:2004zz}, including charmed particles. 
The hard component of the event involves the production of multi$-$jets, modelled using a binomial distribution based on the PYQUEN (PYthia QUENched) energy loss model~\cite{Lokhtin:2014vda}. In this framework, the initial parton spectra are generated using PYTHIA$\_$6.4~\cite{Sjostrand:2006za}, with jet production vertices determined by nuclear geometry that depends on the collision's impact parameter. As the hard partons propagate through the QGP, they experience energy loss influenced by the density and geometry of the soft component, which serves as a static background medium. The collisional energy loss per unit length, $dE^{coll}/dl$, is calculated in the high-momentum transfer limit, representing the incoherent sum of individual scattering events~\cite{Lokhtin:2008xi, Lokhtin:2005px}. Low-momentum transfer contributions are negligible and effectively absorbed by setting $t_{min} \approx  \mu_D^2$, where $\mu_D$ is the Debye screening mass. Meanwhile, the radiative energy loss per unit length, $dE^{rad}/dl$, is evaluated using the BDMS framework~\cite{Lokhtin:2008xi, Baier:1999ds}, which quantifies the strength of multiple scatterings through the transport coefficient $\hat{q}$. This parameter, primarily governed by the initial QGP temperature $T_0$, is used to compute the energy spectrum of coherent, medium-induced gluon radiation~\cite{Lokhtin:2005px}. The final hadronization of hard partons and in-medium emitted gluons according to the Lund string model~\cite{Cimerman:2017lmm} takes place. The number of jets produced in HYDJET++ is directly related to the number of binary NN collisions at a given impact parameter and the integral cross-section of the hard process in NN collisions with the minimum transverse momentum transfer ($p_{T}^{min}$)~\cite{Devi:2024uis}.
\\The calculation of transverse momentum spectra, anisotropic flow, jet-quenching effect, and particle ratios observables for charged particle production in heavy-ion collisions have been successfully done by HYDJET++~\cite{Lokhtin:2009hs, Bravina:2013xla} under the soft and hard component processes.\\ Anisotropic flow is generated from the corresponding spatial eccentricities of the overlap zone of the fireball, which is characterized by the azimuthal spatial anisotropy ($\epsilon(b)$) and momentum anisotropy ($\delta(b)$) parameters~\cite{Devi:2023wih, Bravina:2013xla, Crkovska:2016flo}. These parameters can either be treated independently for each centrality interval or connected through their influence on the elliptic flow coefficient $v_{2}$, whose relationship has been predicted using the hydrodynamical approach~\cite{Lokhtin:2009hs,Devi:2023wih}.\\
In recent studies~\cite{Singh:2023bzm, Devi:2023wih}, the HYDJET++ model has simulated multi$-$strange particle production under a single thermal freeze-out scenario, where all charged particles freeze at the same thermal temperature and $\gamma_s$ ( $\gamma_s=$ 1 shows no strangeness suppression and $\gamma_s$ $\neq$ 1 shows the strangeness suppression). However, other studies~\cite{Waqas:2021bsy, Waqas:2021rmb, ToporPop:2007hb, STAR:2022tfp, Devi:2024dvp, Devi:2024uis} have concluded that thermal$/$chemical freeze$-$out is hadron species$-$dependent, with multi$-$strange hadrons experiencing earlier thermal freeze$-$out due to their smaller hadronic interaction cross$-$sections.
\begin{table}
     \centering
     \begin{tabular}{c|c|c|c}
         Centrality$(\%)$ &  $\langle N_{part}\rangle$ & $\langle N_{coll}\rangle$  &  $\gamma_{s}$  \\
         \hline
         0-5 & 374.5 & 1590 & 1.00  \\
         5-10 &315.5  &1238  & 0.99  \\
         0-10 &345.0  &1414  & -- \\
         10-20 & 242.5 &857.7  &0.975  \\
         20-30 &167.6  &511.0  & 0.95 \\
         30-40 & 111.7 &287.5  & 0.94 \\
         40-50 &70.19  &147.8  & 0.93  \\
         50-60 &40.31  &67.2  & 0.925 \\
         60-80 &14.23  &17.15 & 0.91  \\
       
     \end{tabular}
     \caption{Centrality-dependent values of the number of participants ($\langle N_{\text{part}}\rangle$), number of binary collisions ($\langle N_{\text{coll}}\rangle$), and suppression factor ($\gamma_s$) used in HYDJET++ simulations for Pb+Pb collisions at $\sqrt{s_{\mathrm{NN}}} = 2.76$ TeV.}\label{tab:my_label}
 \end{table}

  \begin{table}
     \centering
     \begin{tabular}{c|c|c|c}
         Centrality$(\%)$ & $\langle N_{part}\rangle$  & $\langle N_{coll}\rangle$  &  $\gamma_{s}$  \\
         \hline
         0-10 & 346.2 & 1540 & 1.00  \\
         10-20 &245.3  &938.7  &0.99   \\
         20-30 &170.1  &558.9  &0.98   \\
         30-40 &114.0 &314.8  &0.95  \\
         40-50 &71.97  &161.6  &0.92   \\
         50-60 &41.64 &73.57 &0.9  \\
         60-70 &21.85  &29.89  &0.89 \\
         70-80 &9.263  &9.811 &0.87  \\
       
     \end{tabular}
     \caption{Centrality-dependent values of the number of participants ($\langle N_{\text{part}}\rangle$), number of binary collisions ($\langle N_{\text{coll}}\rangle$), and suppression factor ($\gamma_s$) used in HYDJET++ simulations for Pb+Pb collisions at $\sqrt{s_{\mathrm{NN}}} = 5.02$ TeV.}\label{tab:my_label1}
 \end{table}
In this article, we have tuned the model on the basis of hadron-species dependent freeze-out parameters like $T_{th}$, $T_{ch}$, and $\gamma_s$. 
We have tuned $\gamma_s$ by matching independently strange/non-strange hadrons particle yield to the experimental data for each centrality interval~\cite{Castorina:2016eyx}. In thermal model calculations, $\gamma_s$ is used as a parameter that indicates the strangeness equilibration as a function of system size and collision energy in p+p and A+A collisions~\cite{Cleymans:2006xj, ALICE:2016fzo, STAR:2006egk, Cleymans:2004pp, Singh:2023bzm, Castorina:2016eyx}. $\gamma_s$ increases with increasing the collision energies $\sqrt{s_{NN}}$ (SPS to LHC then remain constant at LHC)~\cite{ Castorina:2016eyx}. In~\cref {tab:my_label,tab:my_label1}, we have shown the values of $\gamma_s$ given as input; $\langle N_{part}\rangle$, and $\langle N_{coll}\rangle$ for various centrality classes obtained from HYDJET model. By comparing the $p_T$ spectra with ALICE experiment data, we have optimized the value of $\gamma_s$ for both $\sqrt{s_{NN}}$= 2.76 TeV and $\sqrt{s_{NN}}$= 5.02 TeV LHC energies. 
The geometrical parameters obtained from the HYDJET++ model, as calculated using the 2pF (two-parameter Fermi form) MC Glauber model~\cite{Piasecki:2023hoo}, are consistent with the experimental data reported by the ALICE collaborations. The input parameters such as $T_{ch}$ and $T_{th}$, are tunned in such a way that the geometrical quantities remain consistent with the default HYDJET++ model~\cite{Devi:2024dvp, Devi:2024uis}. 
.
\begin{figure*}
    \centering
   \includegraphics[width=0.92\linewidth]{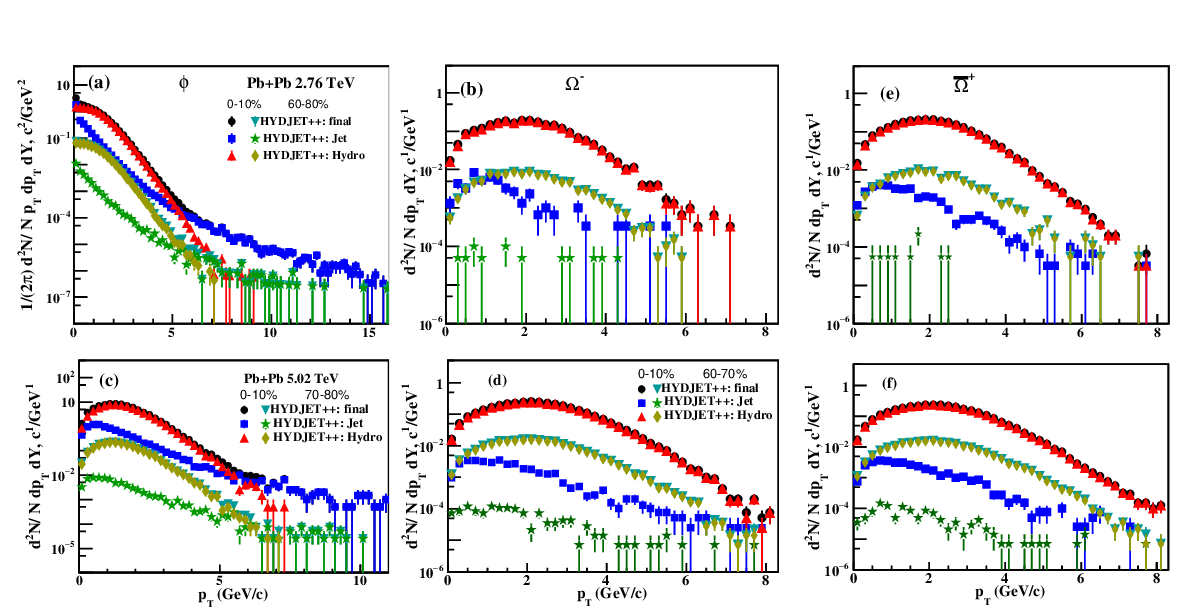}
    \caption{Transverse momentum ($p_T$)-spectra of (a, c) $\phi$, (b, d) $\Omega^{-}$, and (e, f) $\overline{\Omega}^{+}$ in central and peripheral Pb+Pb collisions at $\sqrt{s_{NN}}=2.76$ TeV and $\sqrt{s_{NN}}=5.02$ TeV. 
The spectra obtained from HYDJET++ simulations are shown for the final output, as well as separate contributions from the soft (hydrodynamic) and hard (jet). }
    \label{fig1}
\end{figure*}
\begin{figure*}
    \centering
    \includegraphics[width=0.84\linewidth]{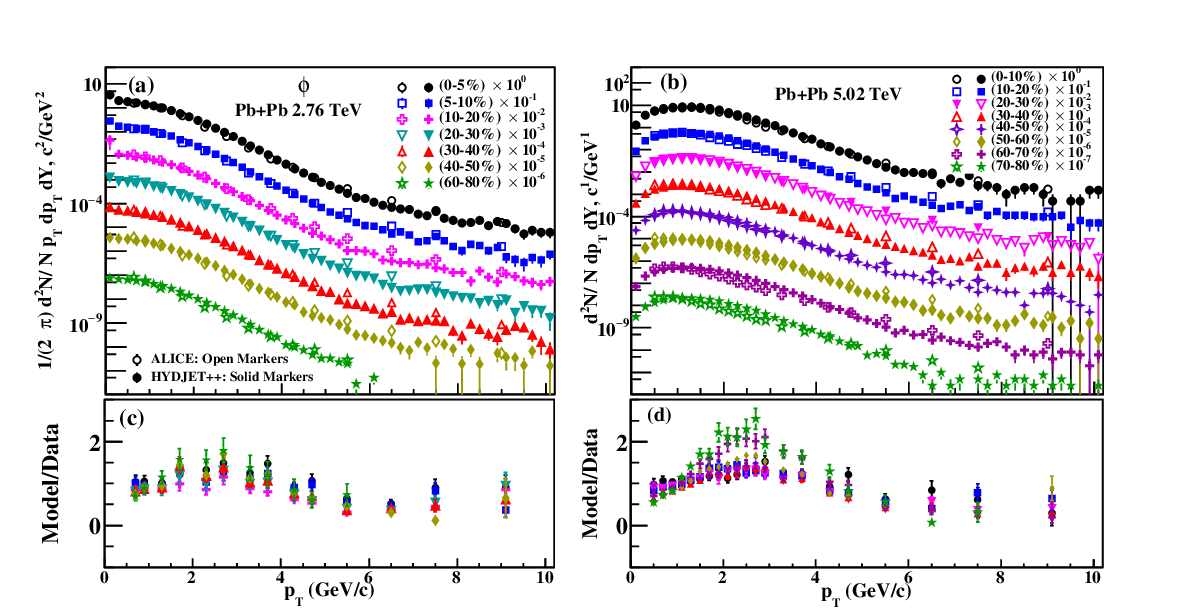}
    \caption{Transverse momentum ($p_T$)-spectra of $\phi$ at (a) $\sqrt{s_{NN}}=2.76$ TeV , (b) $\sqrt{s_{NN}}=5.02$ TeV in Pb+Pb collisions. Different open and solid markers show ALICE experimental results~\cite{ALICE:2017ban, ALICE:2021ptz} and HYDJET++ model results. Panels (c) and (d) show the ratios of model predictions to experimental results.}
    \label{fig2}
\end{figure*}
\begin{figure*}
    \centering
    \includegraphics[width=0.85\linewidth]{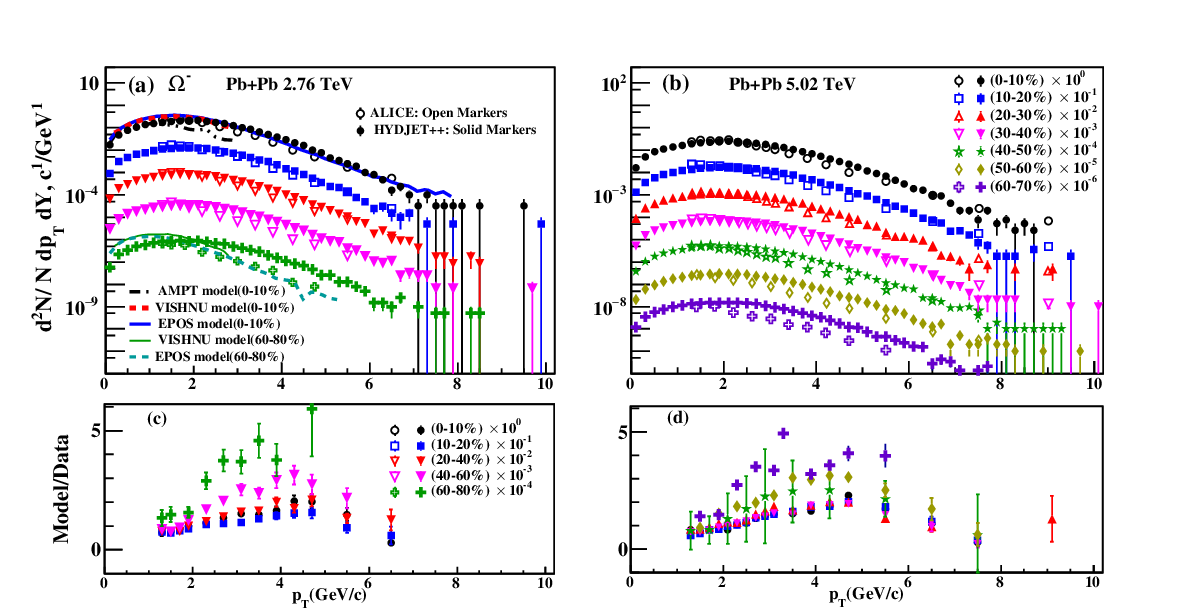}
    \caption{Transverse momentum ($p_T$)-spectra of $\Omega^{-}$ at (a) $\sqrt{s_{NN}}=2.76$ TeV , (b) $\sqrt{s_{NN}}=5.02$ TeV in Pb+Pb collisions. Different open and solid markers show ALICE experimental data~\cite{ALICE:2017ban}, preliminary results~\cite{Kalinak:2017xll}, and HYDJET++ model results. Different lines show the AMPT, VISHNU, and EPOS model results~\cite{ALICE:2013xmt}. Panels (c) and (d) show the ratios of model predictions to experimental data.}
    \label{fig3}
\end{figure*}
\begin{figure*}
    \centering
    \includegraphics[width=0.85\linewidth]{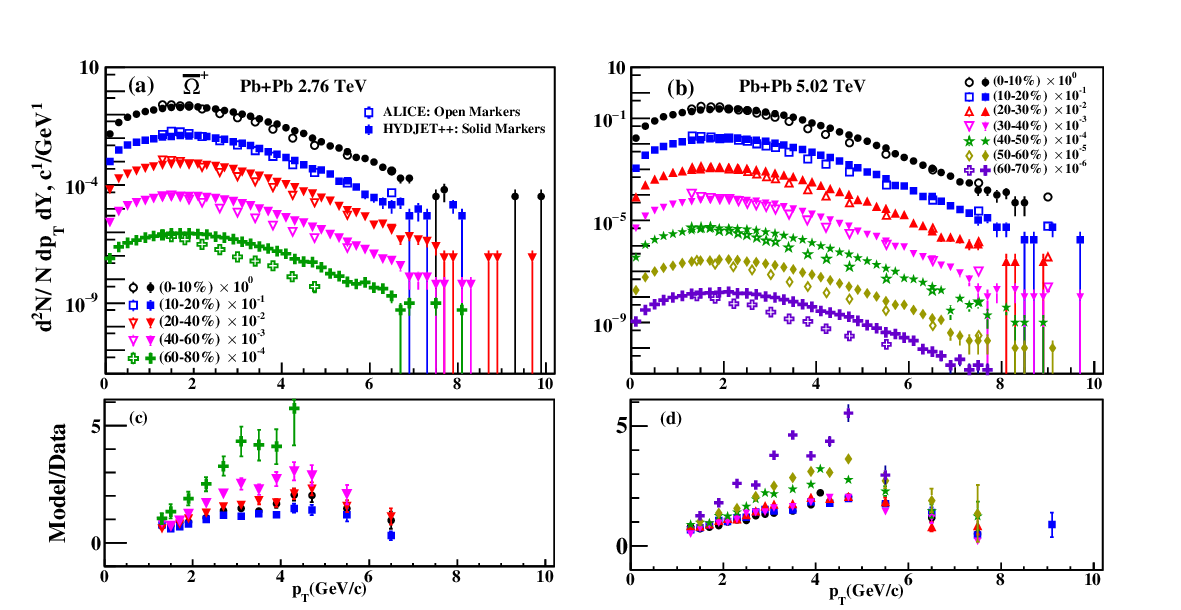}
    \caption{Transverse momentum ($p_T$)-spectra of $\overline{\Omega}^{+}$ at (a) $\sqrt{s_{NN}}=2.76$ TeV, (b) $\sqrt{s_{NN}}=5.02$ TeV in Pb+Pb collisions. Different open and solid markers represent ALICE experimental data~\cite{ALICE:2017ban}, preliminary results~\cite{Kalinak:2017xll}, and HYDJET++ model results. Panels (c) and (d) show the ratios of model predictions to experimental data.}
    \label{fig4}
\end{figure*}
\begin{figure*}
    \centering
    \includegraphics[width=0.95\linewidth]{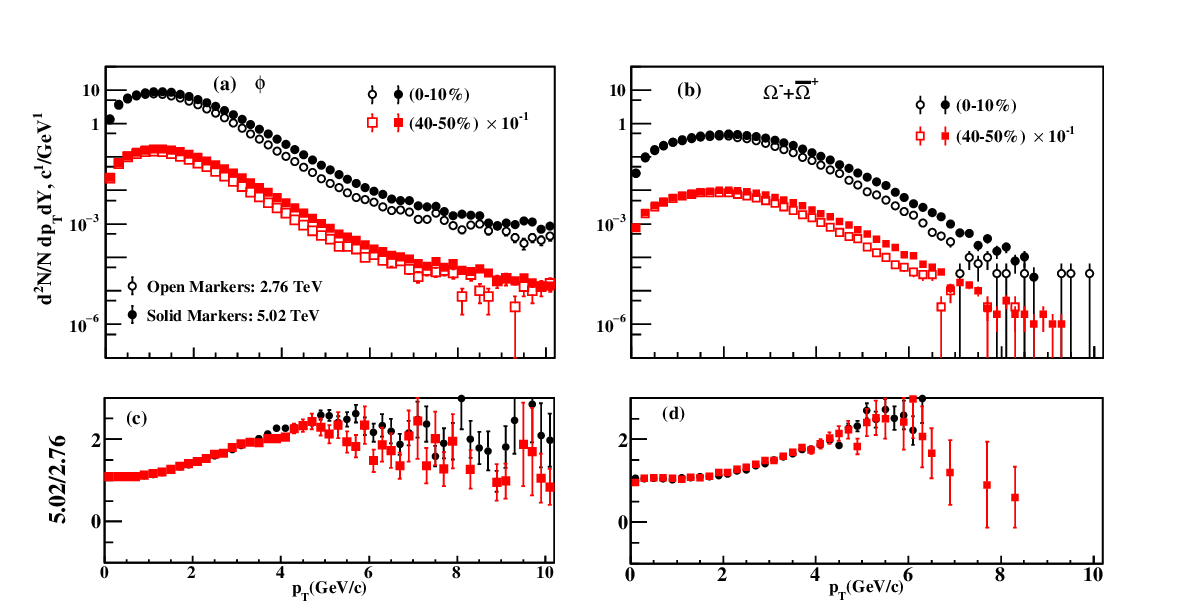}
    \caption{Transverse momentum ($p_T$)-spectra of (a) $\phi$ and (b) $\Omega$ in Pb+Pb collisions at $\sqrt{s_{NN}}$ = 2.76 TeV energy and $\sqrt{s_{NN}}$ = 5.02 TeV energy depicted in the upper panels using different markers for each energy. The corresponding yield ratios between the two energies are shown in the lower panels (c) and (d).}
    \label{fig4.1}
\end{figure*}
\section{Results}\label{Results}
\subsection{$p_{T}-$ spectra}\label{A}
\Cref{fig1,fig2,fig3,fig4} show the $p_{T}$-spectra results of $\phi$-meson and $\Omega$-baryon. \Cref{fig1} shows the $p_T$-spectra in central and peripheral collisions along with soft, and hard contributions separately, and a combination of hard and soft spectra obtained by HYDJET++. 
For $\phi$ and $\Omega$, the bulk of the $p_{T}$ yield have a significant contribution from hydro and jet production at $\sqrt{s_{NN}}=$ 2.76 TeV and $\sqrt{s_{NN}}=$ 5.02 TeV. Moreover, from intermediate to high $p_T$, the particle production is dominated solely by the jet part with a minor contribution from the hydro part. For $\Omega^{-}$ and $\Omega^{+}$, the bulk of the $p_{T}$ yield have a major contribution from hydro (soft) production of hadrons and a minor contribution from hard production in Pb+Pb collisions at $\sqrt{s_{NN}}= $ $2.76$ TeV and $\sqrt{s_{NN}}= $ $5.02$ TeV. It is due to the fact that the $\Omega$ baryon, is less likely to be formed from direct fragmentation, and more likely to result from coalescence~\cite{STAR:2003wqp} (that mechanism is absent in the HYDJET++ model~\cite{Devi:2024uis} and fragmentation dominates at higher energies) within the QGP.\\
In \Cref{fig2,fig3,fig4}, we compared the $p_T$-spectra produced by HYDJET++ with the available ALICE experiment~\cite{ALICE:2017ban, ALICE:2021ptz, ALICE:2019xyr, Kalinak:2017xll} and other models, namely VISHNU, EPOS, and AMPT predictions~\cite{ALICE:2013xmt}. For better visualization, the $p_T$-spectra for different centralities are scaled appropriately with scaling factors mentioned in the legends of the figures. We can see that the HYDJET++ results match quite well with experimental data in central and semi-central collisions for $\phi$ meson, $\Omega^{-}$ and $\Omega^{+}$ baryons.\\
For improved visualization of our results, we have shown the model$-$to$-$data ratios in the lower panels of \Cref{fig2,fig3,fig4}. HYDJET++ substantially overestimates the experimental data in the mid-$p_T$ region for peripheral collisions, particularly for $\Omega^{-}$ and $\Omega^{+}$ baryons at both energies and by a factor of $2.5$ for the $\phi$-meson at $\sqrt{s_{NN}}=5.02$ TeV. This suggests that thermal equilibrium is not achieved for pure strange hadrons in peripheral collisions, mainly due to the absence of coalesence mechanism in the HYDJET model or because of the absence of minijet production in the model~\cite{Bravina:2020sbz}. Furthermore, the comparison of $\Omega^{-}$ baryon $p_T$-spectra with VISHNU and EPOS models in central and peripheral collisions, as well as with the AMPT model in the most central collisions, provides valuable insights. The AMPT model accurately describes the $K^{+}$ spectra but underestimates the $\Lambda$, $\Xi^{-}$, and $\Omega$ spectra, likely due to the absence of strangeness production and annihilation processes in the parton cascade part of the AMPT model~\cite{He:2017tla}. 
The EPOS and VISHNU models overestimate the experimental data at peripheral collisions due to less contribution from feed-down of higher mass resonances as described in ref.~\cite{Devi:2024uis}. \\ 
\Cref{fig4.1}(a), and \Cref{fig4.1}(b) show a comparison in the yields of $\phi$ and $\Omega$ hadrons at $\sqrt{s_{NN}}$ = 2.76 and 5.02 TeV. Experimentally, particle yield should increase with increasing energies and $p_T$ but after a certain $p_T$ value, it should be constant with increasing energies~\cite{ALICE:2021ptz}. Similarly, HYDJET++ results exhibit an increasing trend with increasing energies at a certain $p_T$, however, because of large statistical uncertainties towards high $p_T$, proper physical interpretation cannot be made for the model calculations.
This behaviour gives the information to understand the nuclear modification factors~\cite{ALICE:2021ptz}.\\
\subsection{Nuclear modification factors ($R_{AA}$ and $R_{CP}$)}\label{B}
\begin{figure*}
    \centering
    \includegraphics[width=0.95\linewidth]{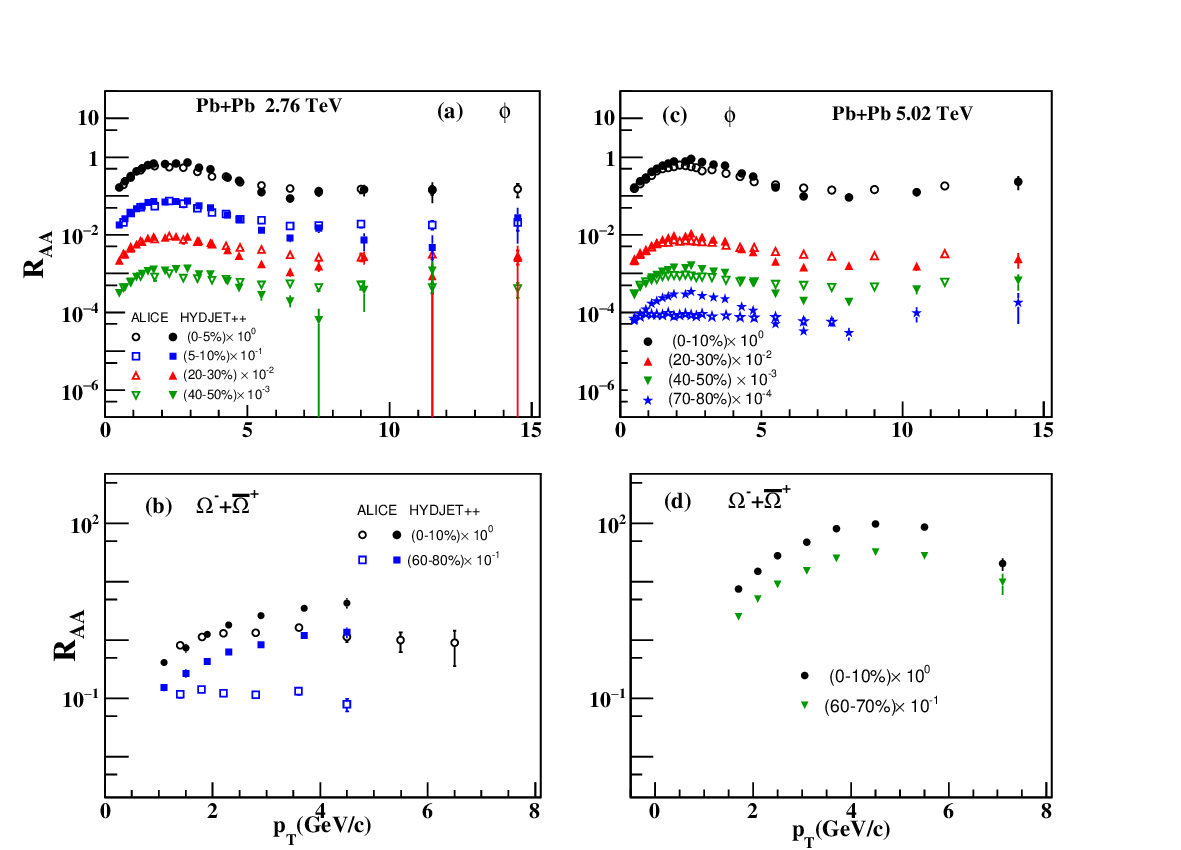}
    \caption{ Nuclear modification factor ($R_{AA}$) of $\phi$ and $\Omega$ at (a,b) $\sqrt{s_{NN}}=2.76$ TeV and (c,d) $\sqrt{s_{NN}}=5.02$ TeV as a function of transverse momentum ($p_{T}$). Different open and solid markers represent ALICE experimental results~\cite{ALICE:2017ban, ALICE:2021ptz,Yin:2013uwa} and HYDJET++ model results, respectively.}
    \label{fig5}
\end{figure*}

\begin{figure*}
    \centering
    \includegraphics[width=0.85\linewidth]{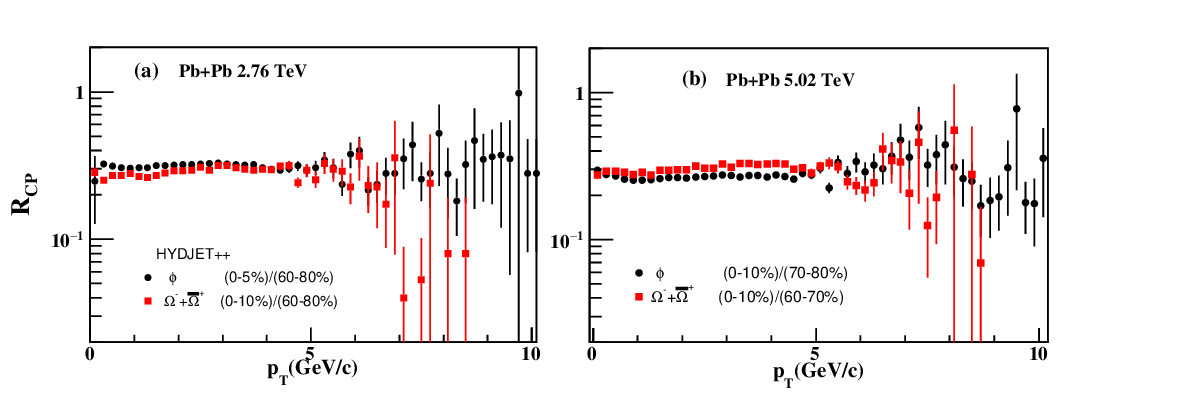}
    \caption{Nuclear modification factor ($R_{CP}$) of $\phi$ and $\Omega$ at (a) $\sqrt{s_{NN}}=2.76$ TeV and (b) $\sqrt{s_{NN}}=5.02$ TeV as a function of transverse momentum ($p_{T}$). The results are obtained from the HYDJET++ model for various centrality intervals, indicated by different markers.}
    \label{fig6}
    \end{figure*}

\Cref{fig5,fig6} show the results of $R_{AA}$ and $R_{CP}$ for $\phi$ meson and $\Omega$ baryon productions in Pb+Pb collisions at $\sqrt{s_{NN}}=2.76$ TeV and $\sqrt{s_{NN}}= 5.02$ TeV energies, respectively at $|y|< 0.5$.  $R_{AA}$ is the ratio of particle yield in A+A collisions with scaling by binary collisions $\langle N_{coll}\rangle$ to particle yield in p+p collisions at the same energies. It gives information about medium effects in heavy-ion collisions relative to p+p collisions. 
On the other hand, $R_{CP}$, which is basically used by theoretical models to understand medium effects, is defined as the ratio of particle yield in central to peripheral collisions with scaling by $\langle N_{coll}\rangle$ at the same energies, collision systems under the same time and conditions~\cite{EuropeanMuon:1983wih, Arneodo:1992wf, Eskola:2016oht, Connors:2017ptx} which reduces other nuclear effects like shadowing, anti-shadowing, etc unlike $R_{AA}$. Due to HYDJET++ model limitations, we can not simulate p+p/p+A collisions results. So, we have used the available ALICE experimental results in p+p collisions~\cite{ALICE:2021ptz} for $R_{AA}$ calculations in both energies.\\
\Cref{fig5}(a) and \Cref{fig5}(c) show that the HYDJET++ model results for $\phi$-meson follow the same trend as experimental results~\cite{ALICE:2017ban, ALICE:2021ptz,Yin:2013uwa} at the entire $p_T$ range for 2.76 TeV and deviate from the experimental results for non-central collisions at 5.02 TeV. The possible explanation for this behaviour in $R_{AA}$ is possibly due to the inability of the model in reproducing the particle spectra in peripheral collisions at 5.02 TeV. \Cref{fig5}(b) shows that the HYDJET++ results for $\Omega^{-}+\overline{\Omega}^{+}$ which follow the same trend as the experimental data but overestimate experimental data at high $p_T$ for $\sqrt{s_{NN}}=2.76$ TeV energy. Similarly, in \Cref{fig5}(d), the results for $\Omega^{-}+\overline{\Omega}^{+}$- baryon follow same trend as 2.76 TeV energy. It is observed that the model qualitatively follows the experimental trend observed for $R_{AA}$ but fails in quantitative description of the experimental results towards high $p_T$ and in peripheral collisions. Both, the model and experimental results show that the $\phi$ mesons experience greater suppression compared to $\Omega$ baryons and exhibit comparable suppression across different energies, whereas $\Omega$ baryons exhibit significant suppression at lower energies.\\
Similarly, \Cref{fig6} shows the HYDJET++ results of $R_{CP}$ for $\phi$-meson and $\Omega$-baryon. It is observed that the $R_{CP}$ is almost independent of the centre-of-mass energy and $p_T$ for the HYDJET model calculations. Based on previous HYDJET++ studies~\cite {Singh:2023bzm, Devi:2023wih, Devi:2024dvp}, as well as the present analysis, the HYDJET++ model demonstrates a good agreement with experimental results for both strange and non-strange hadrons across a wide range of energies from RHIC to LHC.

\begin{figure*}
 \centering
 \includegraphics[width=0.9\linewidth]{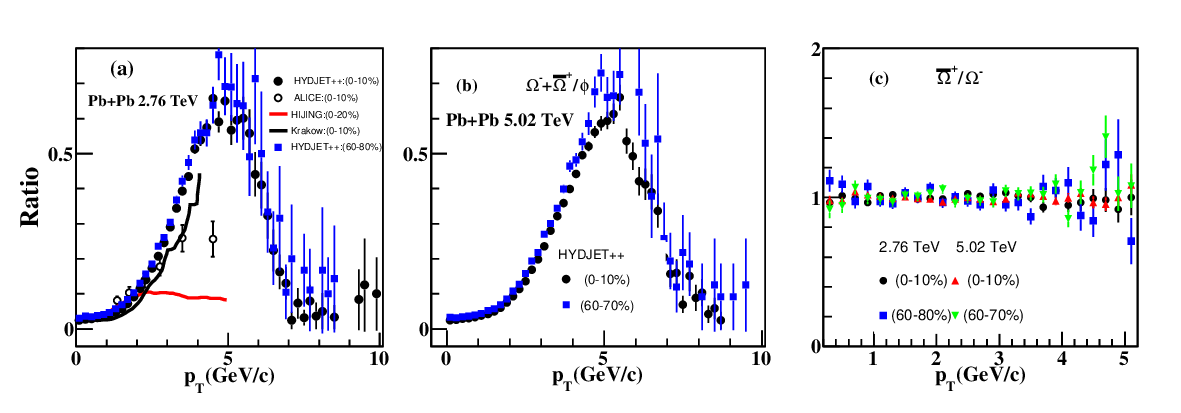}
 \caption{Particle ratio of $\frac{\Omega}{\phi}$ at (a) $\sqrt{s_{NN}}=2.76$ TeV and (b) $\sqrt{s_{NN}}=5.02$ TeV, alongwith the (c) antiparticle-to-particle ratio $\frac{\Omega^{-}}{\overline{\Omega}^{+}}$, as obtained from HYDJET++ simulations. The results are compared with  ALICE experimental results and predictions from the Krakow and HIJING models~\cite{ALICE:2014jbq}.}
    \label{fig7}
\end{figure*}

\begin{figure*}
    \centering
    \includegraphics[width=0.88\linewidth]{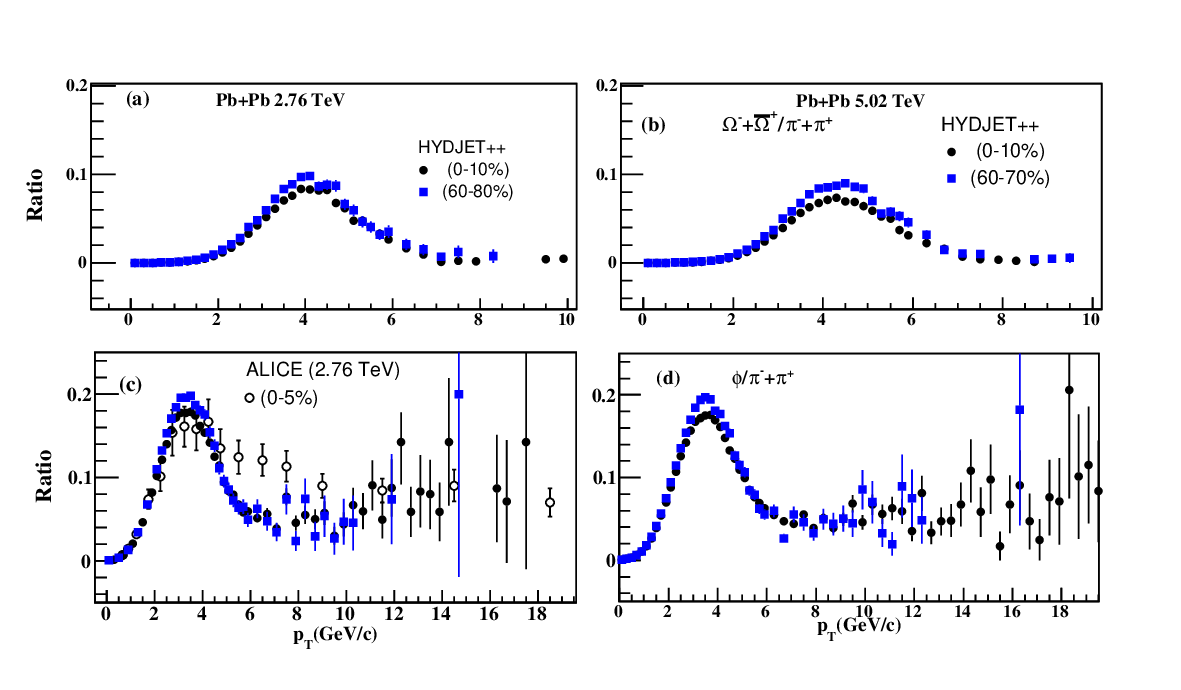}
    \caption{ $\Omega/\pi$ ( in (a), (b)) and $\phi/\pi$ ( in (c), (d)) ratios  as a function of $p_{T}$ for centralities (0-10$\%)$, and (60-80$\%$) at $\sqrt{s_{NN}}= 2.76$ TeV energy, and (0-10$\%$) and (60-70$\%$) at $\sqrt{s_{NN}}= 5.02$ TeV energy. The Open marker shows the ALICE experimental result~\cite{ALICE:2017ban} for $\phi/\pi$ at (0-5$\%$) ( in (c)) centrality.}
    \label{fig8}
\end{figure*}
\subsection{Particle ratio}\label{C}
\Cref{fig7,fig8} show the $\Omega^{-}+\overline{\Omega}^{+}/\phi$, $\overline{\Omega}^{+}/\Omega^{-}$, $\Omega^{-}+\overline{\Omega}^{+}/\pi^{-}+\pi^{+}$ and $\phi/\pi^{-}+\pi^{+}$ ratios as a function of $p_{T}$ in Pb+Pb collisions at $\sqrt{s_{NN}}=2.76$ TeV and $\sqrt{s_{NN}}=5.02$ TeV energies (centrality $(0-10\%)$, $(60-80\%)$, and $(60-70\%)$). These ratios provide helpful information about the chemical properties, such as the chemical equilibrium of the system.\\
Due to small or negligible chemical potential, same-mass particle (i.e., antiparticle-to-particle) ratios are weakly dependent on chemical freeze-out temperature and give less information about the temperature. Different mass-particle ratios are more prominent to check the sensitivity of chemical freeze-out dependence. We have also calculated the different-mass particle ratios such as $\Omega^{-}+\overline{\Omega}^{+}/\phi$, $\Omega^{-}+\overline{\Omega}^{+}/\pi^{-}+\pi^{+}$, and $\phi/\pi^{-}+\pi^{+}$ which serve as a thermometer to probe QGP temperature in heavy-ion collisons~\cite{Cleymans:2006xj}.\\ 
In \Cref{fig7}(a) and \Cref{fig7}(b), $\Omega^{-}+\overline{\Omega}^{+}/\phi$ ratio shows a strong dependence on $p_{T}$ but it is independent of centrality and collision energy.~\cite{ALICE:2014jbq, ALICE:2013mez, PHENIX:2004spo, ALICE:2017ban}. The observed pattern of the $\Omega/\phi$ ratios by HYDJET++ indicate that there is no clear difference in the production of $\Omega^{-}+\overline{\Omega}^{+}$ and $\phi$ particles at various centralities at LHC energies which is similar to the observation of $\phi/K$ ratio at RHIC energies~\cite{PHENIX:2004spo}.\\
We have also compared our findings with available experimental results and Krakow, HIJING/$B\overline{B}$ model results~\cite{ALICE:2014jbq}.
The Krakow model~\cite{Bozek:2012qs} provides predictions for the  $\Omega^{-}+\overline{\Omega}^{+}/\phi$ ratio in the (0-10\%) centrality interval. 
The HYDJET++ simulation shows good agreement with both the Krakow model prediction and ALICE experimental data~\cite{Bozek:2012qs} at low $p_{T}$, and an over prediction at intermediate $p_{T}$ due to non-thermal equilibrium production of $\Omega$ baryons and missing an important mechanism quark coalescence in HYDJET++ which plays a significant role in intermediate $p_T$. Additionally, the available  HIJING/$B\overline{B}$ model~\cite{ToporPop:2011wk, ToporPop:2010qz, Devi:2024uis} result for (0-20\%) centrality interval has also been shown. The model results exhibit a qualitatively similar trend in central collisions at low $p_T$. \Cref{fig7}(c) shows the antiparticle-to-particle ratio. This ratio shows almost independent behaviour with $p_T$ and centrality. 
 Similarly, in \Cref{fig8}, $\Omega^{-}+\overline{\Omega}^{+}/\pi^{-}+\pi^{+}$ and $\phi/\pi^{-}+\pi^{+}$ ratio remain nearly constant with increasing collision energy~\cite{PHENIX:2004spo}. Additionally, the ratio magnitude decreases from lighter to heavier particles with increasing $p_{T}$. It may be because of the high probability of light meson formation rather than heavy strange baryons at low $p_T$. Towards intermediate $p_T$, the ratio of baryons to mesons increases. This is because the particle momentum becomes comparable to or exceeds the mass of strange quarks, significantly enhancing the likelihood of strange baryon formation. At high $p_T$, the fragmentation function of jets is suppressed for strange particles compared to non-strange particles. This results in producing a large number of u and d quarks compared to s quarks and hence the ratios relative to mesons start to decrease at high $p_T$~\cite{Cleymans:2006xj, Singh:2023bzm}. The resulting particle ratios might differ from predicting experimental results at intermediate to high $p_T$ in case of $\Omega^{-}+\overline{\Omega}^{+}/\phi$ and  $\phi/\pi^{-}+\pi^{+}$~\cite{ALICE:2017ban}, perhaps indicating a higher freeze-out temperature or deviating fully from the thermalization picture~\cite{ALICE:2017ban, Cleymans:2006xj}.
\begin{figure*}
    \centering
    \includegraphics[width=0.95\linewidth]{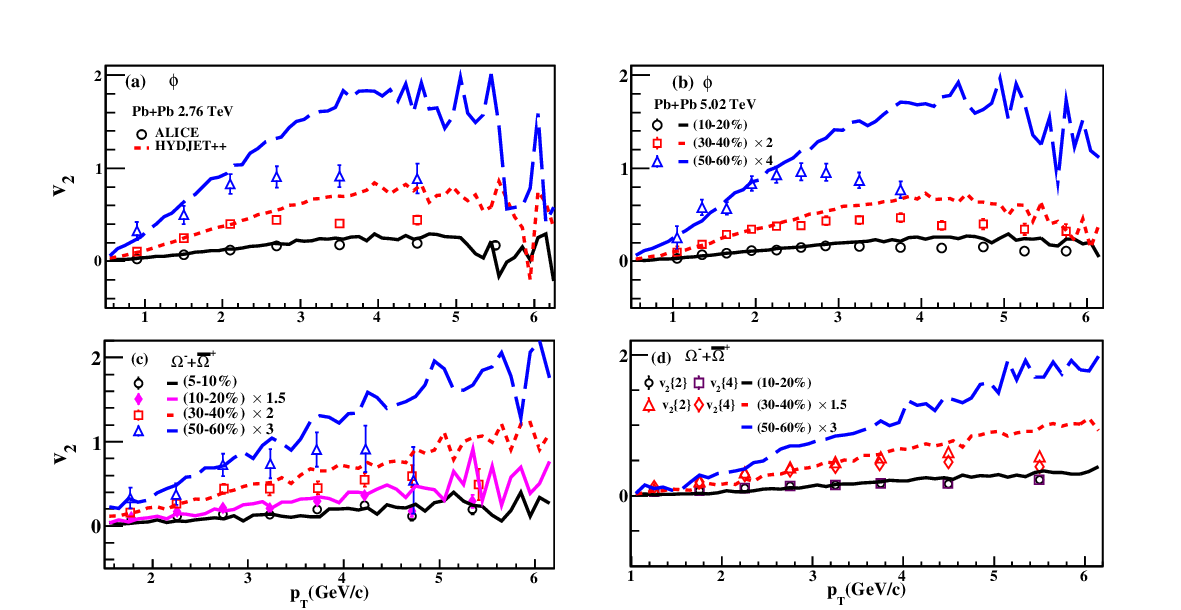}
    \caption{Elliptic flow ($v_2$) of $\phi$ mesons (upper panels: a, b) and $\Omega$ baryons (lower panels: c, d) as a function of transverse momentum ($p_T$). Different Markers show ALICE experimental results~\cite{ALICE:2018yph, ALICE:2014wao}of $v_2$ with various techniques ( 2-particle, 4-particle, and event plane) and lines show HYDJET++ results.}
    \label{fig9}
\end{figure*}
\begin{figure*}
    \centering
    \includegraphics[width=0.75\linewidth]{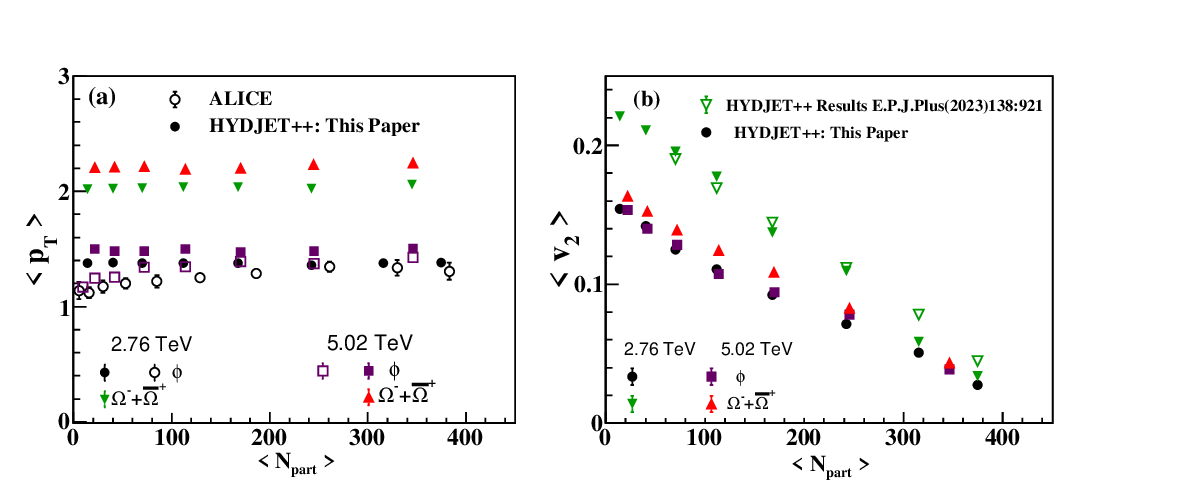}
    \caption{(a) Mean transverse momentum ($\langle p_T\rangle$)~\cite{ALICE:2021ptz} and (b) mean elliptic flow ($\langle v_2\rangle$)~\cite{Devi:2023wih} of $\phi$ and $\Omega$ as a functions of $\left\langle N_{part}\right\rangle$ in Pb+Pb collisions at $\sqrt{s_{NN}}$ = 2.76 TeV and $\sqrt{s_{NN}}$ = 5.02 TeV, respectively.}
    \label{fig10}
\end{figure*}
\begin{figure*}
    \centering
    \includegraphics[width=0.8\linewidth]{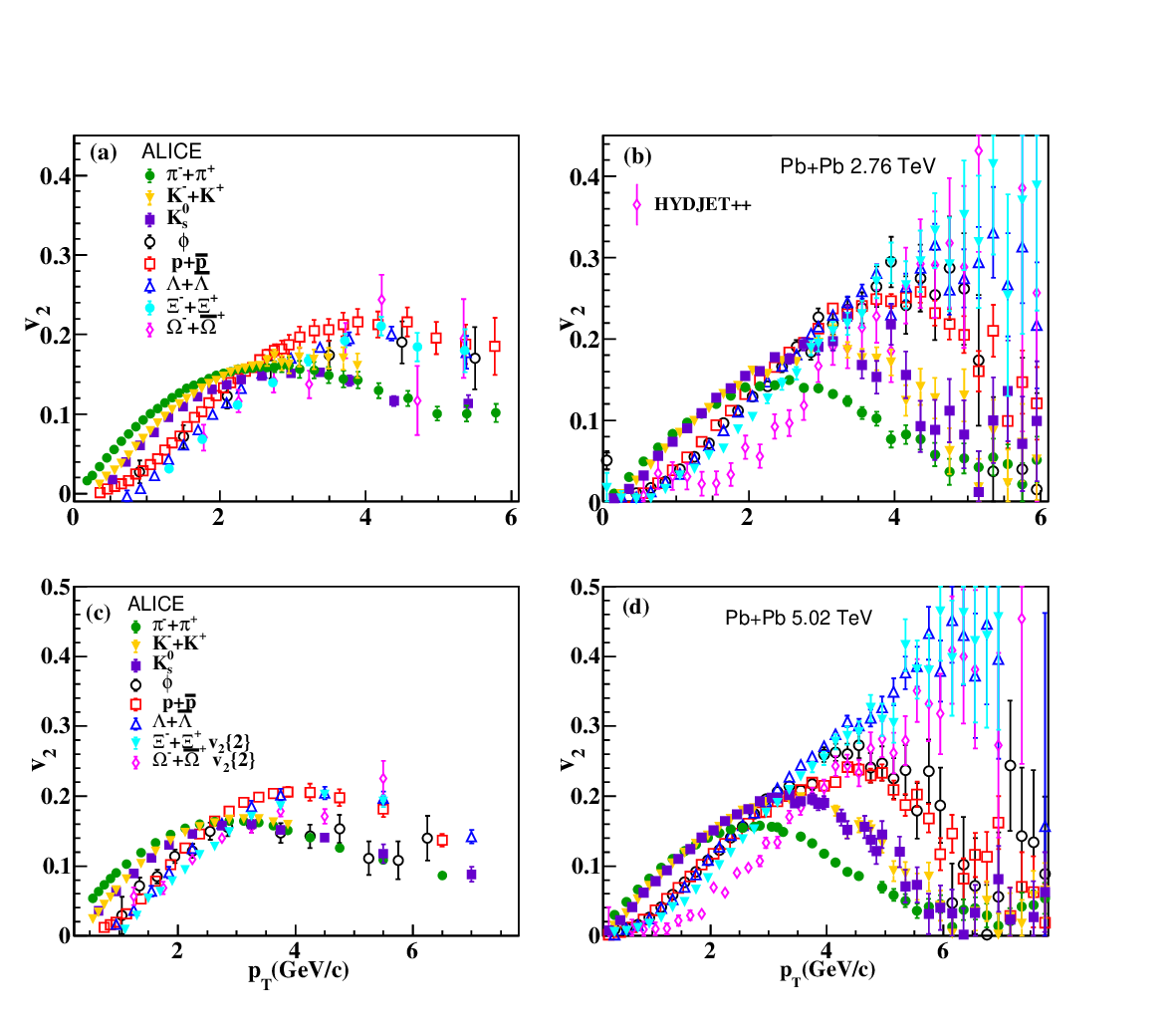}
    \caption{$v_2$ of various particle species for (10-20\%) centrality interval at both energies is shown in separate pads for available ALICE experimental results \cite{ALICE:2018yph, ALICE:2014wao} and HYDJET++ results.}
    \label{fig11}
\end{figure*}
\subsection{Elliptic flow ($v_{2}$)}\label{D}
\Cref{fig9} presents the elliptic flow of $\phi$-meson and $\Omega$-baryons as a function of $p_T$. The elliptic flow shows dependence on $p_T$. The radial flow impels the heavier particles towards higher $p_{T}$ which gets displayed to $p_{T}$-dependent mass ordering of elliptic flow at low $p_{T}$. This tends to decrease the elliptic flow with increased hadron mass~\cite{Devi:2023wih}. Flow occurs due to path-dependent energy loss at intermediate to high $p_T$, offering insights into jet quenching phenomena~\cite{ALICE:2022zks}. For comparison, we have also presented available experimental results in which flow is calculated using different methods such as the event plane, 2-particle, and 4-particle methods~\cite{ALICE:2018yph, ALICE:2014wao}, along with the HYDJET++ reaction plane method. The model accurately reproduces the experimental $v_{2}$ data up to $2-3$ GeV/c in central and mid-central collisions, but overpredicts it at higher $p_{T}$~\cite{Devi:2023wih}. More fluctuations are observed in the $\Omega$-baryon data due to lower statistics, as heavier particle multiplicities are less than those of lighter particles for a given number of events. \\
\Cref{fig10} presents the  $\langle p_T\rangle$ and  $\langle v_{2}\rangle$ of $\phi$-meson and $\Omega$-baryon as a function of $\langle N_{part} \rangle$. The  $\langle p_T\rangle$ in our results exhibits an independent behavior with respect to centrality while $\langle v_{2}\rangle$ shows the strong centrality dependence in $\sqrt{s_{NN}}=2.76$ TeV and $\sqrt{s_{NN}}=5.02$ TeV, respectively. In central collisions, the $\langle p_T \rangle$ of $\phi$-meson from \Cref{fig10}(a) matches well to the experimental data with uncertainty range at both collision energies. However, toward peripheral collisions, the model overpredicts the experimental $\langle p_T \rangle$ data for the $\phi$-meson, indicating that thermal equilibrium might not be achieved for the smaller fireball formed in peripheral collisions. Due to the unavailability of experimental data, we have only presented the model results for $\Omega$'s $\langle p_T\rangle$. The HYDJET++ simulation for the $\Omega$ hadron at both $\sqrt{s_{NN}}=2.76$ TeV and $\sqrt{s_{NN}}=5.02$ TeV are consistent with the recently published results from  HYDJET++(ref.~\cite{Singh:2023bzm}). Similarly, \Cref{fig10}(b) presents the mean $v_2$ results for both energies, obtained using the HYDJET++ model, which is consistent with the findings in a recently published article on HYDJET++~\cite{Devi:2023wih}. The meson-baryon grouping in integrated $p_T$ flow is absent in Pb+Pb collisions from $\sqrt{s_{NN}}=2.76$ TeV to $\sqrt{s_{NN}}=5.02$ TeV.
\subsection{Mass ordering}\label{E}
\Cref{fig11} presents the mass ordering of elliptic flow among $\phi$, $\Lambda$, $p+\overline{p}$ and $\Omega$ in Pb+Pb collisions at $\sqrt{s_{NN}}=2.76$ TeV and $\sqrt{s_{NN}}=5.02$ TeV energies. The available ALICE experimental data~\cite{ALICE:2018yph, ALICE:2014wao} provides the $v_{2}$ mass ordering for $(10–20\%)$ centrality interval at low $p_T$ except for $\phi$, $\Lambda$, and proton. Since these particles have the approximate same mass and different hadronic rescattering contributions (i.e., hadronic state interaction affected by heavy quark contents) that affect their $v_2$ values~\cite{Fries:2010xwl,ALICE:2021ptz}. It is difficult to determine whether this difference is related to the particle's mass or type (whether it is a baryon or a meson). It also suggests that these $v_2$ play a significant role in partonic-to-hadronic dynamics and baryon-to-meson ratio. The HYDJET++ model also follows a similar trend of mass ordering among all the hadron species at both energies.\\

\section{Conclusions}\label{conclusions}
We have studied the particle production of pure multi$-$strange hadrons in Pb+Pb collisions at $\sqrt{s_{NN}}=2.76$ TeV and $\sqrt{s_{NN}}=5.02$ TeV energies, respectively. We have presented the $p_T$-spectra of $\phi$, $\Omega^{-}$, and $\Omega^{+}$ hadrons in both central and peripheral collisions, highlighting the contributions from soft and hard processes. These results provide insights into how the yield contributions from soft and hard processes vary with flavour, particle species, and  energy. We have also compared the $p_T$-spectra at various centrality intervals with available experimental datas and other model results including AMPT, EPOS, and VISHNU. According to our findings, HYDJET++ provides a better description of the experimental data in central collisions at LHC energies for strange particles compared to the VISHNU, AMPT, and EPOS models. However, among these, only the EPOS model accurately describes the $p_T$ spectra of $\Omega^{-}$ baryons at full range of $p_T$ for central to peripheral collisions.\\
Additionally, we compared the yields in central and peripheral collisions at both energies. We observed that at low $p_T< 2$ GeV/c, the charged particle production is expected to be almost the same, while at intermediate $p_T$, there is an increase in charged particle production at higher energies. This provides insights into the nuclear effects relative to p+p/peripheral collisions and enhances our understanding of the behaviour and evolution of the QGP with varying collision energies and system sizes. This effect also contributes to flow dynamics i.e., elliptic flow at high $p_T$. The resulting particle ratios by HYDJET++ might be different from experimental predictions at intermediate to high $p_T$ in case of $\Omega^{-}+\overline{\Omega}^{+}/\phi$ and  $\phi/\pi^{-}+\pi^{+}$, which indicate a deviation from the thermalization picture and make it possible to differentiate between early and late freeze-out. \\
We have also presented the elliptic flow of $\phi$ meson and $\Omega^{-}+\overline{\Omega}^{+}$ baryons and compared our findings with the available ALICE experimental data at both LHC energies. HYDJET++ model suitably describes the experimental data at low $p_T$. At intermediate $p_T$, the model overpredicts the experimental data. The difference between the model calculations and the experimental data increases as we move towards peripheral collisions. This may be because the mechanism involved in addressing the interaction between the medium and jet partons is absent in the HYDJET++ model framework in the present scenarios, particularly with heavier particles. The meson-baryon grouping in integrated $p_T$ flow is absent in $\sqrt{s_{NN}}=2.76$ TeV and $\sqrt{s_{NN}}=5.02$ TeV energies. This behaviour shows that all hadron production is almost the same at high $p_T$ at higher energies. Further,we have given observation of mass ordering in the $v_2$ of strange and non-strange hadrons at low $p_T$ in Pb+Pb collisions for $\sqrt{s_{NN}}=2.76$ TeV and $\sqrt{s_{NN}}=5.02$ TeV energies. 
Both the experimental data and HYDJET++ results show no significant difference between $v_2$ of $\phi$ mesons and protons, which play a substantial role in partonic to hadronic dynamics and baryon to-meson ratios. At intermediate $p_T$, the mass order reverses, and heavier baryons acquire higher flow than mesons.\\
Overall, HYDJET++ delivers deeper insights into the understanding the dynamics of anisotropic flow dynamics, radial expansion effects, and jet-quenching phenomena in high-energy heavy-ion collisions. 

\section*{Acknowledgement}
The authors thank Professor I. P. Lokhtin for useful discussions and for providing expertise in technical issues to understand the HYDJET++ model mechanism. BKS gratefully acknowledges the financial support provided by the BHU Institutions of Eminence (IoE) Grant No. 6031, Govt. of India. GD acknowledges the financial support from UGC under the research fellowship scheme in central universities.

\section*{Data Availability Statement}
The data used to support the findings of this study are included within the article and are cited at relevant places within the text as references. Our finding results will be made available upon request to the corresponding author.


\begin{thebibliography}{5}

\bibitem{STAR:2022tfp}
M.~Abdallah \textit{et al.} [STAR],
Phys. Rev. C \textbf{107} (2023) no.2, 024912
doi:10.1103/PhysRevC.107.024912.
\bibitem{Waqas:2021bsy}
M.~Waqas and G.~X.~Peng,
Entropy \textbf{23} (2021), 488
doi:10.3390/e23040488.

\bibitem{Waqas:2021rmb}
M.~Waqas, G.~X.~Peng and F.~H.~Liu,
J. Phys. G \textbf{48} (2021) no.7, 075108
doi:10.1088/1361-6471/abdd8d.

\bibitem{ToporPop:2007hb}
V.~Topor Pop, M.~Gyulassy, J.~Barrette, C.~Gale, S.~Jeon and R.~Bellwied,
Phys. Rev. C \textbf{75} (2007), 014904
doi:10.1103/PhysRevC.75.014904.

\bibitem{Devi:2024uis}
G.~Devi, A.~Singh and B.~K.~Singh,
J. Phys. G \textbf{51} (2024) no.9, 095203
doi:10.1088/1361-6471/ad63c0.

\bibitem{STAR:2015gge}
L.~Adamczyk \textit{et al.} [STAR],
Phys. Rev. Lett. \textbf{116} (2016) no.6, 062301
doi:10.1103/PhysRevLett.116.062301.

\bibitem{Bugaev:2016voh}
K.~A.~Bugaev, D.~R.~Oliinychenko, V.~V.~Sagun, A.~I.~Ivanytskyi, J.~Cleymans, E.~S.~Mironchuk, E.~G.~Nikonov, A.~V.~Taranenko and G.~M.~Zinovjev,
Ukr. J. Phys. \textbf{61} (2016), 659-673.

\bibitem{NA49:2004irs}
C.~Alt \textit{et al.} [NA49],
Phys. Rev. Lett. \textbf{94} (2005), 192301
doi:10.1103/PhysRevLett.94.192301.

\bibitem{STAR:2015vvs}
L.~Adamczyk \textit{et al.} [STAR],
Phys. Rev. C \textbf{93} (2016) no.2, 021903
doi:10.1103/PhysRevC.93.021903.

\bibitem{PHENIX:2013kod}
A.~Adare \textit{et al.} [PHENIX],
Phys. Rev. C \textbf{88} (2013) no.2, 024906
doi:10.1103/PhysRevC.88.024906.


\bibitem{STAR:2019bjj}
J.~Adam \textit{et al.} [STAR],
Phys. Rev. C \textbf{102} (2020) no.3, 034909
doi:10.1103/PhysRevC.102.034909.

\bibitem{Ye:2017ooc}
Y.~J.~Ye, J.~H.~Chen, Y.~G.~Ma, S.~Zhang and C.~Zhong,
Chin. Phys. C \textbf{41} (2017) no.8, 084101
doi:10.1088/1674-1137/41/8/084101.


\bibitem{Devi:2024dvp}
G.~Devi, A.~Singh and B.~K.~Singh,
DAE Symp. Nucl. Phys. \textbf{67} (2024), 1061-1062.

\bibitem{ALICE:2021lsv}
S.~Acharya \textit{et al.} [ALICE],
Eur. Phys. J. C \textbf{81} (2021) no.7, 584
doi:10.1140/epjc/s10052-021-09304-4.
\bibitem{Feng:2022sdh}
Y.~T.~Feng, F.~L.~Shao and J.~Song,
Phys. Rev. C \textbf{106} (2022) no.3, 034910
doi:10.1103/PhysRevC.106.034910.
\bibitem{PHENIX:2004spo}
S.~S.~Adler \textit{et al.} [PHENIX],
Phys. Rev. C \textbf{72} (2005), 014903
doi:10.1103/PhysRevC.72.014903.
\bibitem{Singh:2023bzm}
A.~Singh, P.~K.~Srivastava, G.~Devi and B.~K.~Singh,
Phys. Rev. C \textbf{107} (2023) no.2, 024906
doi:10.1103/PhysRevC.107.024906.

\bibitem{ALICE:2021ptz}
S.~Acharya \textit{et al.} [ALICE],
Phys. Rev. C \textbf{106} (2022) no.3, 034907.

\bibitem{Braun-Munzinger:2003htr}
P.~Braun-Munzinger, J.~Stachel and C.~Wetterich,
Phys. Lett. B \textbf{596} (2004), 61-69
doi:10.1016/j.physletb.2004.05.081.

\bibitem{NA49:2008goy}
C.~Alt \textit{et al.} [NA49],
Phys. Rev. C \textbf{78} (2008), 044907
doi:10.1103/PhysRevC.78.044907.

\bibitem{ALICE:2014wao}
B.~B.~Abelev \textit{et al.} [ALICE],
JHEP \textbf{06} (2015), 190 doi:10.1007/JHEP06(2015)190.
\bibitem{Devi:2023wih}
G.~Devi, A.~Singh, S.~Pandey and B.~K.~Singh,
Eur. Phys. J. Plus \textbf{138} (2023) no.10, 921.
\bibitem{ALICE:2018yph}
S.~Acharya \textit{et al.} [ALICE],
JHEP \textbf{09} (2018), 006.
\bibitem{STAR:2003wqp}
J.~Adams \textit{et al.} [STAR],
Phys. Rev. Lett. \textbf{92} (2004), 052302
doi:10.1103/PhysRevLett.92.052302.
\bibitem{Ortiz:2017cul}
A.~Ortiz and O.~V\'azquez,
Phys. Rev. C \textbf{97} (2018) no.1, 014910
doi:10.1103/PhysRevC.97.014910.
\bibitem{Christiansen:2013hya}
P.~Christiansen, K.~Tywoniuk and V.~Vislavicius,
Phys. Rev. C \textbf{89} (2014) no.3, 034912
doi:10.1103/PhysRevC.89.034912.





\bibitem{Lokhtin:2008xi}
I.~P.~Lokhtin, L.~V.~Malinina, S.~V.~Petrushanko, A.~M.~Snigirev, I.~Arsene and K.~Tywoniuk,
Comput. Phys. Commun. \textbf{180} (2009), 779-799.

\bibitem{Lokhtin:2009hs} I.~P.~Lokhtin, L.~V.~Malinina, S.~V.~Petrushanko, A.~M.~Snigirev, I.~Arsene and K.~Tywoniuk, Nonlin. Phenom. Complex Syst. \textbf{12} (2009), 348-355.

\bibitem{Lokhtin:2005px} I.~P.~Lokhtin and A.~M.~Snigirev,
Eur. Phys. J. C \textbf{45} (2006), 211-217.

\bibitem{Zhu:2016qiv} X.~Zhu,
Adv. High Energy Phys. \textbf{2016} (2016), 4236492
doi:10.1155/2016/4236492.

\bibitem{He:2017tla}
Y.~He and Z.~W.~Lin,
Phys. Rev. C \textbf{96} (2017) no.1, 014910
doi:10.1103/PhysRevC.96.014910.

\bibitem{Lin:2021mdn}
Z.~W.~Lin and L.~Zheng,
Nucl. Sci. Tech. \textbf{32} (2021) no.10, 113
doi:10.1007/s41365-021-00944-5.

\bibitem{Zhu:2015dfa}
X.~Zhu, F.~Meng, H.~Song and Y.~X.~Liu,
Phys. Rev. C \textbf{91} (2015) no.3, 034904
doi:10.1103/PhysRevC.91.034904.

\bibitem{Pierog:2013ria}
T.~Pierog, I.~Karpenko, J.~M.~Katzy, E.~Yatsenko and K.~Werner,
Phys. Rev. C \textbf{92} (2015) no.3, 034906
doi:10.1103/PhysRevC.92.034906.

\bibitem{Werner:2007bf}
K.~Werner,
Phys. Rev. Lett. \textbf{98} (2007), 152301
doi:10.1103/PhysRevLett.98.152301.
\bibitem{Bozek:2012qs}
P.~Bozek and I.~Wyskiel-Piekarska,
Phys. Rev. C \textbf{85} (2012), 064915
doi:10.1103/PhysRevC.85.064915.

\bibitem{ToporPop:2010qz}
V.~Topor Pop, M.~Gyulassy, J.~Barrette, C.~Gale and A.~Warburton,
Phys. Rev. C \textbf{83} (2011), 024902
doi:10.1103/PhysRevC.83.024902.

\bibitem{ToporPop:2011wk}
V.~Topor Pop, M.~Gyulassy, J.~Barrette and C.~Gale,
Phys. Rev. C \textbf{84} (2011), 044909
doi:10.1103/PhysRevC.84.044909. 
\bibitem{Lin:2004en}
Z.~W.~Lin, C.~M.~Ko, B.~A.~Li, B.~Zhang and S.~Pal,
Phys. Rev. C \textbf{72} (2005), 064901
doi:10.1103/PhysRevC.72.064901.

\bibitem{Cimerman:2017lmm}
J.~Cimerman, B.~Tomasik, M.~Csanad and S.~Lokos,
Eur. Phys. J. A \textbf{53} (2017) no.8, 161
doi:10.1140/epja/i2017-12349-7.
\bibitem{Cleymans:2004pp}
J.~Cleymans, B.~Kampfer, M.~Kaneta, S.~Wheaton and N.~Xu,
Phys. Rev. C \textbf{71} (2005), 054901
doi:10.1103/PhysRevC.71.054901.
\bibitem{Cleymans:2006xj}
J.~Cleymans, I.~Kraus, H.~Oeschler, K.~Redlich and S.~Wheaton,
Phys. Rev. C \textbf{74} (2006), 034903
doi:10.1103/PhysRevC.74.034903.
\bibitem{ALICE:2014jbq}
B.~B.~Abelev \textit{et al.} [ALICE],
Phys. Rev. C \textbf{91} (2015), 024609.
\bibitem{ALICE:2013xmt}
B.~B.~Abelev \textit{et al.} [ALICE],
Phys. Lett. B \textbf{728} (2014), 216-227
[erratum: Phys. Lett. B \textbf{734} (2014), 409-410.
\bibitem{Amelin:2007ic} N.~S.~Amelin, R.~Lednicky, I.~P.~Lokhtin, L.~V.~Malinina, A.~M.~Snigirev, I.~A.~Karpenko, Y.~M.~Sinyukov, I.~Arsene and L.~Bravina, Phys. Rev. C \textbf{77} (2008), 014903. 
\bibitem{Crkovska:2016flo}
J.~Crkovska, J.~Bielcik, L.~Bravina, B.~H.~Brusheim Johansson, E.~Zabrodin, G.~Eyyubova, V.~L.~Korotkikh, I.~P.~Lokhtin, L.~V.~Malinina and S.~V.~Petrushanko, \textit{et al.}
Phys. Rev. C \textbf{95} (2017) no.1, 014910
doi:10.1103/PhysRevC.95.014910.
\bibitem{Piasecki:2023hoo}
K.~Piasecki and P.~Piotrowski,
Eur. Phys. J. A \textbf{59} (2023) no.11, 272
doi:10.1140/epja/s10050-023-01182-6.
\bibitem{Torrieri:2004zz}
G.~Torrieri, S.~Steinke, W.~Broniowski, W.~Florkowski, J.~Letessier and J.~Rafelski,
Comput. Phys. Commun. \textbf{167} (2005), 229-251
doi:10.1016/j.cpc.2005.01.004.

\bibitem{Lokhtin:2014vda}
I.~P.~Lokhtin, A.~A.~Alkin and A.~M.~Snigirev,
Eur. Phys. J. C \textbf{75} (2015) no.9, 452
doi:10.1140/epjc/s10052-015-3594-3.
\bibitem{Sjostrand:2006za}
T.~Sjostrand, S.~Mrenna and P.~Z.~Skands,
JHEP \textbf{05} (2006), 026
doi:10.1088/1126-6708/2006/05/026.
\bibitem{Baier:1999ds}
R.~Baier, Y.~L.~Dokshitzer, A.~H.~Mueller and D.~Schiff,
Phys. Rev. C \textbf{60} (1999), 064902
doi:10.1103/PhysRevC.60.064902.
\bibitem{Bravina:2013xla}
L.~V.~Bravina, B.~H.~Brusheim Johansson, G.~K.~Eyyubova, V.~L.~Korotkikh, I.~P.~Lokhtin, L.~V.~Malinina, S.~V.~Petrushanko, A.~M.~Snigirev and E.~E.~Zabrodin,
Eur. Phys. J. C \textbf{74} (2014) no.3, 2807
doi:10.1140/epjc/s10052-014-2807-5.
\bibitem{Castorina:2016eyx}
P.~Castorina, S.~Plumari and H.~Satz,
Int. J. Mod. Phys. E \textbf{25} (2016) no.08, 1650058
doi:10.1142/S0218301316500580.
\bibitem{ALICE:2016fzo}
J.~Adam \textit{et al.} [ALICE],
Nature Phys. \textbf{13} (2017), 535-539
doi:10.1038/nphys4111.

\bibitem{STAR:2006egk}
J.~Adams \textit{et al.} [STAR],
Phys. Rev. Lett. \textbf{98} (2007), 062301
doi:10.1103/PhysRevLett.98.062301.

\bibitem{ALICE:2017ban}
J.~Adam \textit{et al.} [ALICE],
Phys. Rev. C \textbf{95} (2017) no.6, 064606.

\bibitem{ALICE:2019xyr}
S.~Acharya \textit{et al.} [ALICE],
Phys. Lett. B \textbf{802} (2020), 135225.

\bibitem{Kalinak:2017xll}
P.~Kalinak [ALICE],
PoS \textbf{EPS-HEP2017} (2017), 168.
\bibitem{Bravina:2020sbz}
L.~V.~Bravina, G.~K.~Eyyubova, V.~L.~Korotkikh, I.~P.~Lokhtin, S.~V.~Petrushanko, A.~M.~Snigirev and E.~E.~Zabrodin,
Phys. Rev. C \textbf{103} (2021) no.3, 034905
doi:10.1103/PhysRevC.103.034905.
\bibitem{EuropeanMuon:1983wih}
J.~J.~Aubert \textit{et al.} [European Muon],
Phys. Lett. B \textbf{123} (1983), 275-278
doi:10.1016/0370-2693(83)90437-9.

\bibitem{Eskola:2016oht}
K.~J.~Eskola, P.~Paakkinen, H.~Paukkunen and C.~A.~Salgado,
Eur. Phys. J. C \textbf{77} (2017) no.3, 163
doi:10.1140/epjc/s10052-017-4725-9.

\bibitem{Arneodo:1992wf}
M.~Arneodo,
Phys. Rept. \textbf{240} (1994), 301-393
doi:10.1016/0370-1573(94)90048-5.

\bibitem{Connors:2017ptx}
M.~Connors, C.~Nattrass, R.~Reed and S.~Salur,
Rev. Mod. Phys. \textbf{90} (2018), 025005
doi:10.1103/RevModPhys.90.025005.




\bibitem{Yin:2013uwa}
Z.~Yin [ALICE],
Int. J. Mod. Phys. Conf. Ser. \textbf{29} (2014), 1460228
doi:10.1142/S2010194514602282.
\bibitem{ALICE:2013mez}
B.~Abelev \textit{et al.} [ALICE],
Phys. Rev. C \textbf{88} (2013), 044910
doi:10.1103/PhysRevC.88.044910.

\bibitem{ALICE:2022zks}
S.~Acharya \textit{et al.} [ALICE],
JHEP \textbf{05} (2023), 243
doi:10.1007/JHEP05(2023)243.






\bibitem{Fries:2010xwl}
R.~J.~Fries,
PoS \textbf{CERP2010} (2010), 008
doi:10.22323/1.111.0008.

\end{thebibliography}
\end{document}